\documentclass[aip, pop, reprint]{revtex4-1}
\usepackage{graphicx} 
\usepackage{dcolumn}
\usepackage{bm}
\usepackage{amssymb}
\usepackage{amsmath}
\usepackage{wasysym}
\usepackage{color}
\usepackage{hyperref}
\usepackage{dcolumn}
\usepackage{float}
\usepackage{flushend}
\usepackage{multirow}
\usepackage{cancel}
\hypersetup{
colorlinks,
citecolor=blue,
filecolor=blue,
linkcolor=blue,
urlcolor=blue}
\newcommand{\la}{\left\langle}
\newcommand{\ra}{\right\rangle}
\newcommand{\be}{\begin{equation}}
\newcommand{\ee}{\end{equation}}
\newcommand{\bse}{\begin{subequations}}
\newcommand{\ese}{\end{subequations}}
\newcommand{\bea}{\begin{eqnarray}}
\newcommand{\eea}{\end{eqnarray}}
\newcommand{\ba}{\begin{array}}
\newcommand{\ea}{\end{array}}

\begin{document}
\title{Turbulent drag reduction in magnetohydrodynamic and quasi-static magnetohydrodynamic turbulence}
\author{Mahendra K. Verma}
\email{mkv@iitk.ac.in}
\affiliation {Department of Physics, Indian Institute of Technology, Kanpur, India 208016}

\author{Shadab Alam}
\email{shadab@iitk.ac.in}
\affiliation {Department of Mechanical Engineering, Indian Institute of Technology, Kanpur, India 208016}

\author{Soumyadeep Chatterjee}
\email{soumyade@iitk.ac.in}
\affiliation {Department of Physics, Indian Institute of Technology, Kanpur, India 208016}

%\keywords{Turbulence $|$ Turbulent drag reduction $|$Energy transfers $|$ Energy flux $|$ Magnetohydrodynamic turbulence} 

\begin{abstract} 
In hydrodynamic turbulence, the kinetic energy injected at large scales cascades to the inertial range, leading to  a constant kinetic energy flux.  In contrast, in magnetohydrodynamic (MHD) turbulence, a fraction of kinetic energy is transferred to the magnetic energy.  Consequently, for the same kinetic energy injection rate, the kinetic energy flux in MHD turbulence is reduced compared to its hydrodynamic counterpart.  This leads to relative weakening of the nonlinear term ($\la | ({\bf u}\cdot \nabla) {\bf u} | \ra$, where ${\bf u}$ is the velocity field) and  turbulent drag, but strengthening of the velocity field in  MHD turbulence.   We verify the above using shell model simulations of hydrodynamic and MHD turbulence.  Quasi-static MHD turbulence too exhibits turbulent drag reduction similar to MHD turbulence.

\end{abstract}

%An important factor in the drag reduction is the energy transfer from the velocity field to the polymers.
\maketitle
 
% If your first paragraph (i.e. with the \dropcap) contains a list environment (quote, quotation, theorem, definition, enumerate, itemize...), the line after the list may have some extra indentation. If this is the case, add \parshape=0 to the end of the list environment.

\section{Introduction}
Frictional force in  a turbulent flow   is proportional to  square of the flow velocity~\cite{Davidson:book:Turbulence,Sagaut:book}.  This  steep  dependence of the frictional force or {\em turbulent drag} on the velocity makes it a major  challenge in aerospace and automobile industry, as well as for flow engineering.  Past experiments and numerical simulations  report turbulent drag reduction in a dilute solution with  polymers (see ~\citet{Tabor:EPL1986,deGennes:book:Intro,Sreenivasan:JFM2000,Benzi:PD2010,Benzi:ARCMP2018}, and references therein) and in solutions with bubbles and surfactants~\cite{Lvov:PRL2005}.   
Despite  many experimental and theoretical attempts, we are far from consensus on the mechanism behind this phenomenon.    Researchers attribute the following factors for the drag reduction: viscoelasticity, nonlinear interactions between the polymer and the velocity field, interactions at the boundary layers, anisotropic stress, etc.~\cite{Tabor:EPL1986,deGennes:book:Intro,Sreenivasan:JFM2000,Benzi:PD2010,Benzi:ARCMP2018}.  Both, bulk and boundary layer dynamics may play a significant role in drag reduction. Yet, some researchers argue that the contributions from the bulk probably dominates that from the boundary layer~\cite{Sreenivasan:JFM2000}. { In the present paper we study drag reduction in the bulk region of  magnetohydrodynamics (MHD) and quasi-static magnetohydrodynamics (QSMHD) turbulence, but not for the bluff bodies in such flows.  }

In this paper,  using the energy fluxes we show that MHD and QSMHD turbulence  exhibit turbulent drag reduction.  In particular, we demonstrate that an inclusion of magnetic field in a turbulent flow leads to reductions in kinetic energy cascade rate,  the nonlinear term $\la | ({\bf u}\cdot \nabla) {\bf u} | \ra$ (where ${\bf u}$ is the velocity field, and $\la . \ra$ represents averaged value), and turbulent drag.  In particular, we describe the turbulent drag reduction from the perspectives of energy transfers and energy flux in the bulk flow.

In a turbulent flow forced at large scales, the injected  kinetic energy  cascades to intermediate scale and then to small scales, where the energy flux is dissipated by viscous force. In pure hydrodynamic turbulence, the energy injection rate, kinetic energy flux, and the viscous dissipation are equal~\cite{Kolmogorov:DANS1941Dissipation,Kolmogorov:DANS1941Structure,Davidson:book:Turbulence,Sagaut:book}, and the turbulent drag is proportional to the kinetic energy flux.   In MHD turbulence and  QSMHD turbulence, a part of  kinetic energy  is converted to  magnetic energy, as illustrated by a large body of works on energy transfer computations in MHD and QSMHD turbulence~\cite{Cowling:book, Dar:PD2001,Alexakis:PRE2005,Mininni:PRE2005S2S,Debliquy:PP2005,Kumar:EPL2014,Verma:ROPP2017,Alexakis:PR2018}.  The above energy transfers lead to a reduction in kinetic energy flux and the magnitude of the nonlinear term, and hence a reduction in turbulent drag in MHD and QSMHD turbulence.  

Reduction in kinetic energy flux also  takes place in polymeric turbulence and in bubbly turbulence.  A common feature among these systems is that the kinetic energy is transferred to the elastic energy associated with the magnetic field, polymers, and bubbles.  Here, magnetic field in MHD acts like a taut string; polymers  as springs; and bubbles as elastic spheres.   { It is important to note that MHD turbulence is more complex than hydrodynamic turbulence~\cite{Sagaut:book, Onsagar:Nouvo1949_SH,Eyink:RMP2006,Eyink:arxiv2018}}.

 To demonstrate the above we performed numerical simulations of  shell models of hydrodynamic and MHD turbulence with identical kinetic energy injection rates.  We observe that the kinetic energy flux and the nonlinear term $\la | ({\bf u}\cdot \nabla) {\bf u} | \ra$  for
	MHD turbulence are lower than the corresponding quantities for hydrodynamic turbulence, but the flow speeds for MHD turbulence is larger than that for hydrodynamic counterpart.  We remark that turbulent drag reduction in MHD and QSMHD turbulence would be important for astrophysical and engineering applications.  { For example, magnetic field is imposed to suppress velocity fluctuations in crystal growth and plate rolling\cite{Bojarevics:book_chapter, Bochkarev:book_chapter, Chudinovskij:book_chapter}.  We will describe the applications to dynamo later in the paper.}

The outline of the paper is as follows.  In Sections~\ref{sec:Hydro} and \ref{sec:MHD} we relate turbulent drag reduction in hydrodynamic and hydromagnetic turbulence to energy fluxes.  In Section~\ref{sec:shellmodel} we perform numerical simulations of hydrodynamic and MHD turbulence using shell models and verify that the turbulent drag is reduced in MHD turbulence.  Sections~\ref{sec:QSMHD} and \ref{sec:polymer} cover drag reduction issues in QSMHD and polymeric turbulence respectively.  We conclude in Sec.~\ref{sec:conclusions}.
%%%%

\section{Turbulent drag in hydrodynamic turbulence}
\label{sec:Hydro}

In this section we briefly describe the  energy flux in Kolmogorov's theory of hydrodynamic turbulence, and relate it to the turbulent drag.  The equations for incompressible hydrodynamics are
\bea
\frac{\partial{\bf u}}{\partial t} + ({\bf u}\cdot\nabla){\bf u}
& = & -\nabla({p}/{\rho}) +  \nu\nabla^2 {\bf u}   +  {\bf F}_\mathrm{ext},  \label{eq:U}  \\
\nabla \cdot {\bf u}  & = & 0, \label{eq:incompress}
\eea
where ${\bf u}, p$ are respectively the velocity  and pressure fields; $\rho$ is the density which is assumed to be unity;   $\nu$ is  the kinematic viscosity;      $ {\bf F}_\mathrm{ext}$ is the external force employed at large scales that helps  maintain a steady state. 

The multiscale energy transfer and dynamics are conveniently described using Fourier modes.  The external force $ {\bf F}_\mathrm{ext}$  injects kinetic energy at large scales. For a wavenumber ${\bf k}$, the kinetic energy injection rate is
\be
\mathcal{F}_\mathrm{ext}({\bf k})  =    \Re[ {\bf F}_\mathrm{ext}({\bf k}) \cdot {\bf u}^*({\bf k}) ],
\label{eq:Fmathcal_def}
\ee
where $\Re$ stands for the real part of the argument.   We denote the total kinetic energy injection rate using $ \epsilon_\mathrm{inj}$. That is,
\be
   \int_0^{k_f}  d{\bf k} \mathcal{F}_\mathrm{ext}({\bf k}) \approx \epsilon_\mathrm{inj}.
\ee
This injected kinetic energy cascades to larger wavenumbers as  kinetic flux $ \Pi_u(k) $.  In the inertial range, $ \Pi_u(k) = \epsilon_\mathrm{inj}$ due to the absence of external force and negligible viscous dissipation~\cite{Kolmogorov:DANS1941Dissipation, Davidson:book:Turbulence, Frisch:book, Sagaut:book}.

\begin{figure}%[tbhp]
	\centering
	\includegraphics[width=1\linewidth]{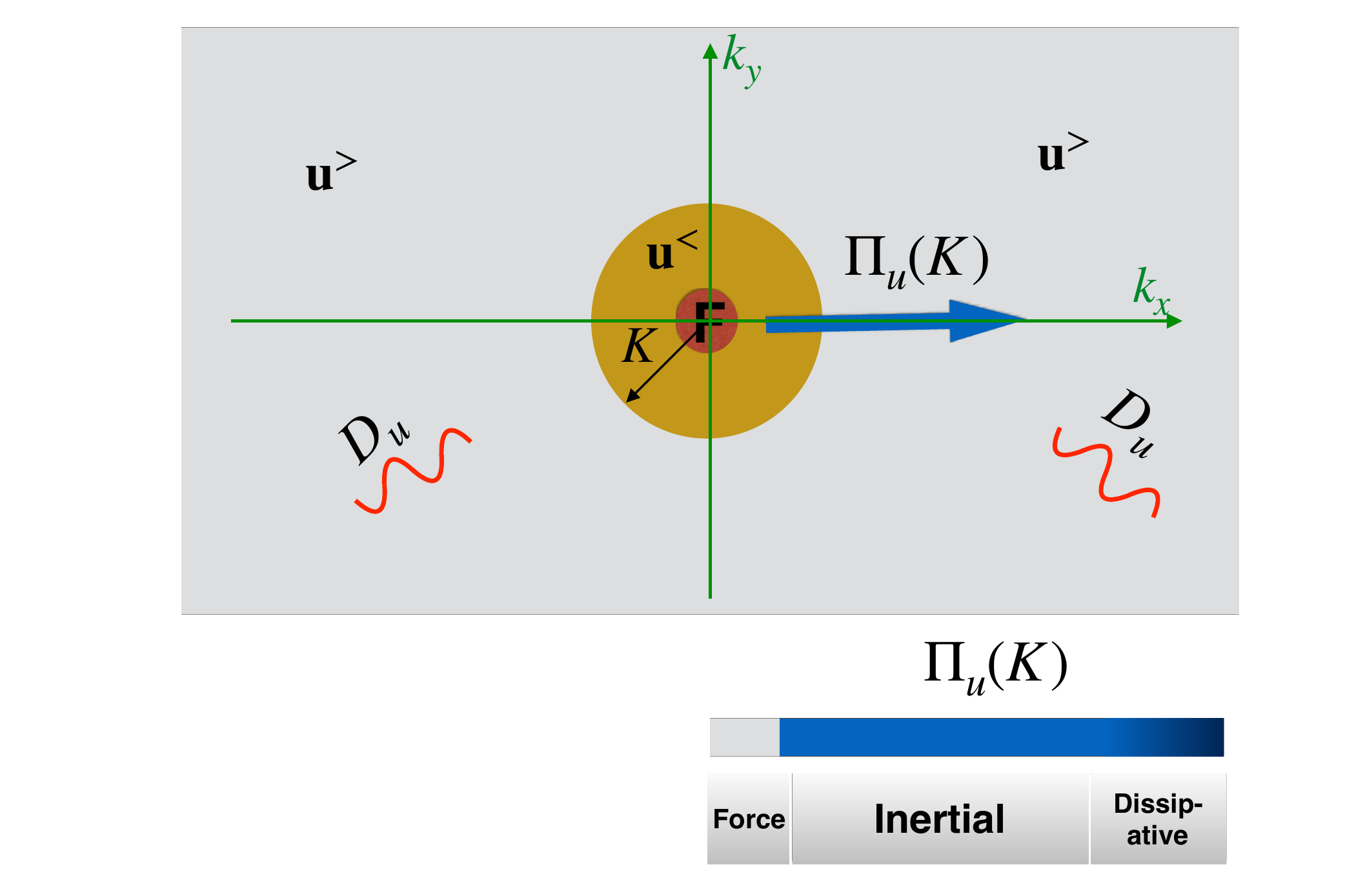}
	\caption{(color online)   In hydrodynamic turbulence, the energy supplied by the external force ${\bf F}_\mathrm{ext}$ at small wavenumbers (red sphere) cascades to the inertial range, leading to a constant kinetic energy flux $\Pi_u(k)$ in the inertial range. This flux is dissipated at small scales by viscosity.    In the colorbar, the blue color represents constant  $\Pi_u(k)$, while the black color represents depleted $\Pi_u(k)$. }
	\label{fig:flux}
\end{figure}

This energy flux is dissipated in the dissipative range, and the total viscous dissipation rate is given by
\be
\epsilon_u = \int d{\bf k} D_u({\bf k}) = \int d{\bf k}   2 \nu k^2 E_u({\bf k}),
\ee
where
\be
E_u({\bf k}) = \frac{1}{2} |{\bf u(k)}|^2
\ee
is modal kinetic energy, and
\be
D_u({\bf k})   = 2 \nu k^2 E_u({\bf k}) 
\label{eq:Du_def}
\ee
is the modal viscous dissipation rate.  Based on energetic arguments, we conclude that~\cite{Kolmogorov:DANS1941Dissipation, Frisch:book, Sagaut:book}
\be
\Pi_u(k) \approx     \epsilon_\mathrm{inj} \approx \epsilon_u \approx  \frac{U^3}{d} ,
\ee
where $d$ is the large length scale of the system, and $U$ is the large scale velocity.  In this paper we take $U$ as root mean square (rms) velocity.   We illustrate the kinetic energy flux and the viscous dissipation  in Fig.~\ref{fig:flux}. 

For the large scales, under a steady state, Eq.~(\ref{eq:U}) yields the average turbulent drag $F_{D,\mathrm{HD}}$ as
 \be
 \la F_{D,\mathrm{HD}} \ra_\mathrm{LS} \approx \la  |{\bf (u \cdot \nabla) u}|  \ra_\mathrm{LS} \approx \la   F_\mathrm{ext}   \ra ,
 \ee
 where $\la . \ra_\mathrm{LS}$ represents ensemble averaging over the large scales.  In terms of energy flux,
 \be
 U F_{D,\mathrm{HD}}   \approx \Pi_u \approx \frac{U^3}{d} \approx \epsilon_\mathrm{inj},
 \label{eq:FD_HD_flux}
 \ee
 or
  \be
 F_{D,\mathrm{HD}}   \approx \frac{\Pi_u}{U} \approx    \frac{U^2}{d}.
 \label{eq:FD_HD}
 \ee
 Note that the viscous dissipation can be ignored at  large scales.     

In the next section we will describe how the energy transfers from the velocity field to the magnetic field leads to a drag reduction in MHD turbulence.

\section{Drag reduction in MHD turbulence in terms of  energy flux}
\label{sec:MHD}
 The equations for incompressible MHD turbulence are  
\bea
\frac{\partial{\bf u}}{\partial t} + ({\bf u}\cdot\nabla){\bf u}
& = & -\nabla({p}/{\rho}) +  \nu\nabla^2 {\bf u} + {\bf F}_u({\bf B,B}) +  {\bf F}_\mathrm{ext},  \label{eq:U_MHD}  \nonumber \\ \\ 
\frac{\partial{\bf B}}{\partial t} + ({\bf u}\cdot\nabla){{\bf B}}
& = &   \eta \nabla^2 {{\bf B}} +  {\bf F}_B({\bf B,u}),   \label{eq:W} \\
\nabla \cdot {\bf u}  & = & 0,\\
\nabla \cdot {\bf B}  &= & 0, \label{eq:incompress}
 \eea
 where ${\bf u,B}$ are respectively the velocity and magnetic fields; $p$ is the total (thermal + magnetic) pressure, $\rho$ is the density which is assumed to be unity;   $\nu$ is  the kinematic viscosity;  $ \eta$ is the magnetic diffusivity;   $ {\bf F}_\mathrm{ext}$ is the external force employed  at large scales; and 
 \bea
 {\bf F}_u & = &  {\bf (B \cdot \nabla) B}, \\
 {\bf F}_B & = & {\bf (B \cdot \nabla) u}
 \eea represent respectively the Lorentz force and the stretching of the magnetic field by the velocity field.  Such interactions lead to energy exchanges among the field variables.  In the above equations, the magnetic field ${\bf B}$ is in units of velocity, which is achieved by using ${\bf B} = {\bf B}_\mathrm{cgs}/\sqrt{4\pi \rho}$. Here, ${\bf B}_\mathrm{cgs}$ is the magnetic field in CGS units.

\begin{figure}%[tbhp]
  	\centering
  	\includegraphics[width=1\linewidth]{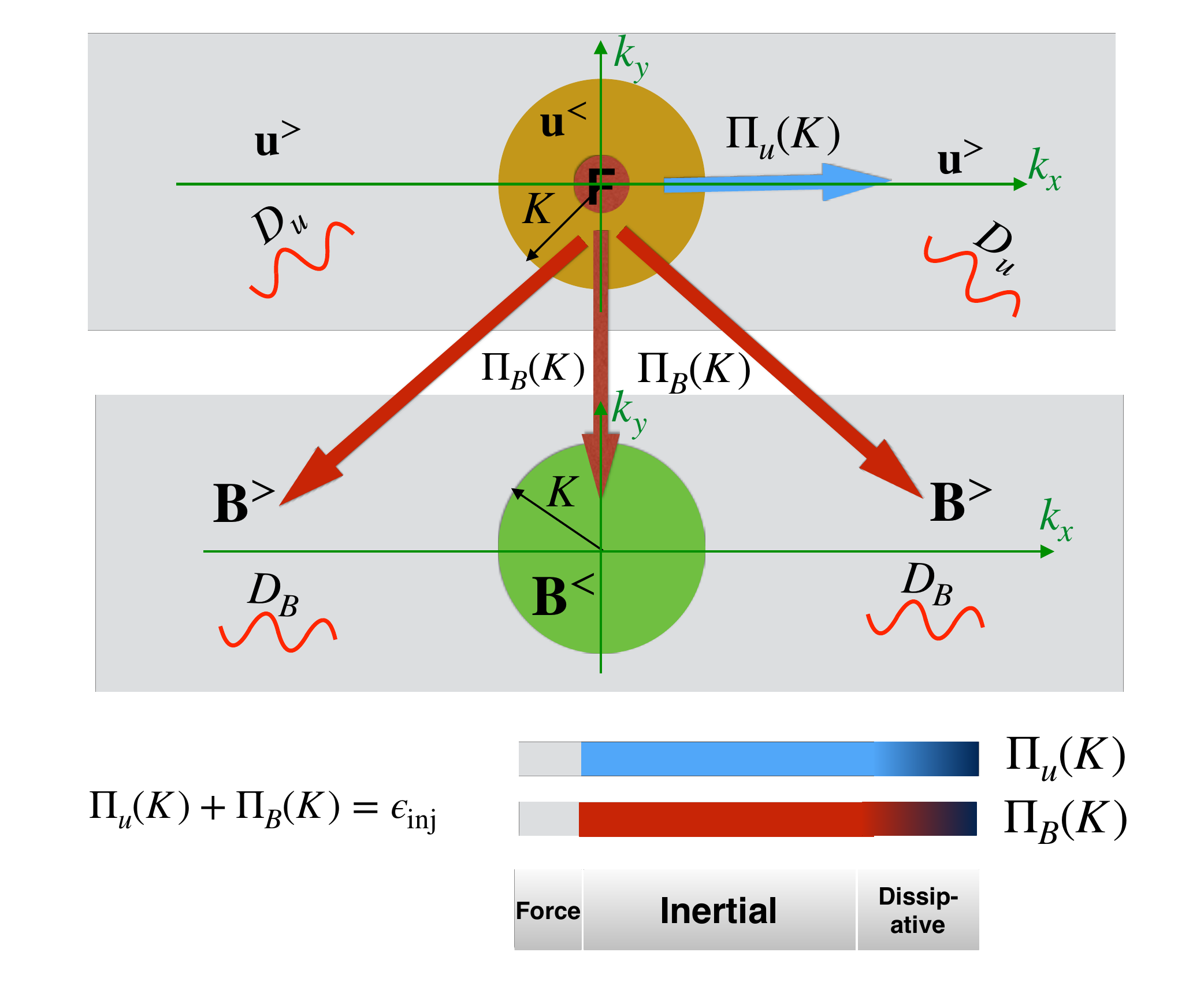}
  	\caption{(color online)  In MHD turbulence, a fraction of  kinetic energy flux is transferred to the magnetic field.  $\Pi_u(K)$ is the kinetic energy flux for the velocity wavenumber sphere of radius $K$ (yellow sphere), and $\Pi_B(K)$ is the net energy transfer from ${\bf u}$ modes inside the sphere to all the ${\bf B}$ modes.   The   external force injects kinetic energy  into the small red sphere with the rate of $\epsilon_\mathrm{inj}$.  The energy fluxes are dissipated with  the dissipation rates $D_u$ and $D_B$.  In the inertial range, $\Pi_u(K) + \Pi_B(K)  \approx \epsilon_\mathrm{inj}$.    In the colorbars, the light-blue color represents a constant  kinetic energy flux $\Pi_u(k)$, while the red color represents $\Pi_B(K)$.  The black color in the dissipation range represents depleted fluxes. Light-blue colored $\Pi_u(k)$ for MHD turbulence is lower than corresponding dark-blue colored $\Pi_u(k)$ for hydrodynamic turbulence. }
  	\label{fig:MHD_flux}
  \end{figure}

Computation of multiscale energy transfers are quite convenient in spectral or Fourier  space, in which the evolution equation for the modal kinetic energy $E_u({\bf k}) = |{\bf u(k)}|^2/2$ is~\citep{Kraichnan:JFM1959,Frisch:book, Verma:PR2004,Davidson:book:Turbulence,Sagaut:book,Verma:book:BDF}   
 \bea
  \frac{d}{dt} E_u(\mathbf{k})  & = & T_{u}({\bf k})  +  \mathcal{F}_u({\bf k})  + \mathcal{F}_\mathrm{ext}({\bf k})-D_u(\mathbf{k}),
 \label{eq:MHD_ET:Eu_dot_Fext} 
\eea
where 
 \bea
 T_u({\bf k}) & = & \sum_{\bf p} \Im \left[ {\bf  \{  k \cdot u(q) \} \{ u(p) \cdot u^*(k) \} } \right] , \\
 \mathcal{F}_u({\bf k}) & =  & \Re[ {\bf F}_u({\bf k}) \cdot {\bf u}^*({\bf k})  ] \nonumber \\
 & = & \sum_{\bf p} - \Im \left[ {\bf  \{  k \cdot B(q) \} \{ B(p) \cdot u^*(k) \} } \right]
 \eea
 with $\Re, \Im$  representing the real and imaginary parts respectively, and ${\bf q=k-p}$.   Note that $\mathcal{F}_\mathrm{ext}({\bf k})$ and $D_u(\mathbf{k}) $ have been defined in Eqs.~(\ref{eq:Fmathcal_def}) and (\ref{eq:Du_def}) respectively. In this paper we do not discuss the energetics of ${\bf B}$ field because the turbulent drag reduction is related to the energy fluxes associated with the velocity field.  When we sum Eq.~(\ref{eq:MHD_ET:Eu_dot_Fext}) over all the modes in the wavenumber sphere of radius $K$, we obtain the following equation~\cite{Davidson:book:Turbulence,Sagaut:book,Verma:book:BDF}:
 \bea
 \frac{d}{dt} \sum_{k \le K}  E_u({\bf k})   &= &  \sum_{k \le K} T_u({\bf k}) + \sum_{k \le K}\mathcal{F}_u({\bf k})  \nonumber \\
 &&   + \sum_{k \le K}\mathcal{F}_\mathrm{ext}({\bf k}) - \sum_{k \le K} D_u({\bf k}).
 \label{eq:ET:Pi_k0_from_Ek}
  \eea  
 The first term $\sum_{k\le K} T_u({\bf k}) $ is the net energy transfer from the modes outside the sphere to the modes inside the sphere due to the nonlinear term $ ({\bf u}\cdot\nabla){\bf u}$.  It is also negative of the kinetic energy flux for the sphere because
  \bea
  \Pi_u(K) = -\sum_{k \le K} T_u({\bf k}).
  \eea
The second term $\sum_{k\le K} \mathcal{F}_u({\bf k}) $ represents the total energy transfer rate to the velocity modes inside the sphere from all the magnetic modes~\cite{Dar:PD2001,Verma:PR2004}.  Since
   \bea
  \Pi_B(K) = - \sum_{k \le K}\mathcal{F}_u({\bf k}),
  \eea
  the second term  is  negative of the net energy flux from the velocity modes inside the sphere to all the magnetic modes.  The third term $\sum_{k\le K} \mathcal{F}_\mathrm{ext}({\bf k}) $ is the net energy injected by the external force ${\bf F}_\mathrm{ext}$ (represented by the red sphere of Fig.~\ref{fig:MHD_flux}). The last term $\sum_{k\le K} D_u({\bf k})$ is the total viscous dissipation rate inside the sphere.   See Figure~\ref{fig:MHD_flux} for an illustration of the above quantities.    
 
 For a  wavenumber sphere of radius $K$, under a steady state ($d   \sum_{k \le K}   E_u({\bf k}) /dt =0$),  the kinetic energy injected  by ${\bf F}_\mathrm{ext}$ is lost  to the two fluxes ($\Pi_u$, $ \Pi_B$) and the total viscous dissipation rate. That is, 
 \be
 \Pi_u(K) + \Pi_B(K) + \sum_{k \le K} D_u({\bf k}) = \epsilon_\mathrm{inj}.
 \ee
 In the inertial range where $D_u({\bf k}) \approx 0$, we obtain
  \be
 \Pi_u(K) + \Pi_B(K)  \approx \epsilon_\mathrm{inj}.
 \label{eq:Pi_sum}
 \ee
  Following similar lines of arguments as in the previous section, we estimate the turbulent drag in MHD turbulence using
 \be
 \la F_{D,\mathrm{MHD}} \ra \approx \la  |{\bf (u \cdot \nabla) u}|  \ra_\mathrm{LS} \approx    \frac{\Pi_u}{U}   .
 \ee
  Using Eq.~(\ref{eq:Pi_sum}) we deduce that
  \be
   \la F_{D,\mathrm{MHD}} \ra \approx   \frac{ \epsilon_\mathrm{inj}- \Pi_B}{U}.
   \ee

 Researchers have studied the energy fluxes $\Pi_u$ and $\Pi_B$ in detail for various combinations of parameters---forcing functions, boundary condition, $\nu$ and $\eta$ (or their ratio $\mathrm{Pm} = \nu/\eta$, which is called the {\em magnetic Prandtl number}).    For example, Mininni et al.~\citep{Mininni:ApJ2005} computed the fluxes  $\Pi_u$ and $\Pi_B$ using numerical simulations, and observed that   $\Pi_B > 0$, and  hence
\be
\Pi_{u,\mathrm{MHD}} <  \Pi_{u,\mathrm{HD}}.
  \label{eq:Pi_MHD_reduced}
\ee   
That is, the kinetic energy flux in MHD turbulence is lower than the corresponding flux in hydrodynamic turbulence (without magnetic field).  
Using numerical simulations,  Debliquy et al.~\citep{Debliquy:PP2005}, Kumar et al.~\citep{Kumar:EPL2014},  Verma and Kumar~\citep{Verma:JoT2016} arrived at a similar conclusion.  In particular, using a shell model, Verma and Kumar~\citep{Verma:JoT2016} simulated MHD turbulence for Pm = 1  and showed that  in the inertial range, $\Pi_u \approx 0.06 \pm 0.02$ and $\Pi_B \approx 0.93 \pm 0.02$; this result indicates a drastic reduction of kinetic energy flux in MHD turbulence.  Note that, $\Pi_B$,  the energy transferred  from the velocity field to the magnetic field  is responsible  for the enhancement of magnetic field in astrophysical dynamos (e.g., in planets, stars, and galaxies)~\cite{Moffatt:book}.    Laboratory  experiments~(see \citet{Monchaux:PF2009} and references therein) too exhibit similar energy transfers.  Based on these observations on the energy fluxes, we deduce that
  \be
F_{D,\mathrm{MHD}} <  F_{D,\mathrm{HD}},
\label{eq:F_D_reduced}
\ee 
 or turbulent drag is reduced in MHD turbulence.     Quantitatively, it will be more appropriate to compute drag-reduction coefficients:
\be
c_1 = \frac{ \Pi_{u} }{U^3/d};~~c_2 = \frac{ \la |{\bf (u \cdot \nabla) u}| \ra }{U^2/d}.
\label{eq:c1_c2}
\ee
Note that $c_1, c_2 \approx 1$ for hydrodynamic turbulence, and they are expected to be lower than unity for MHD turbulence.

The aforementioned reduction in the energy flux and the nonlinear term $\la |{\bf (u \cdot \nabla) u}| \ra$  lead to {stronger large-scale velocity ($U$) in MHD turbulence than in the corresponding hydrodynamic turbulence.}    Note that strong and coherent $U$ yields a diminished    $\la |{\bf (u \cdot \nabla) u}| \ra$ and kinetic energy flux.  This is because the kinetic energy flux depends on $U$ as well {as} on the phase relations between the velocity Fourier modes~\cite{Kraichnan:JFM1959, Verma:PR2004}.  {We illustrate the above based on the numerical simulations of  \citet{Yadav:PRE2012}.  They showed that in a subcritical dynamo transition, $U$ increases abruptly when the magnetic field becomes nonzero at the onset of dynamo (see Fig. 4 of \citet{Yadav:PRE2012}). Note that in the dynamo simulations of \citet{Yadav:PRE2012}, the flow  exhibits turbulent behaviour even though the Reynolds number is not very large.  Hence, the energy fluxes are expected to be significant in these situations. }  Contrast these features with those in laminar regime in which the the Lorentz force  increases the  drag. 

 Though the past results on energy fluxes provide strong credence towards drag reduction in MHD turbulence, we need a clear-cut demonstration of the same.  Towards this objective, we performed simulations of shell models for hydrodynamic and MHD turbulence with same kinetic energy injection rates.  We show that the the kinetic energy flux and  $\la |{\bf (u \cdot \nabla) u}| \ra$  are reduced for MHD turbulence, but $U$ for MHD  turbulence is larger than that for hydrodynamic turbulence. We will report these results in the next section.

\section{Numerical verification using shell model}
\label{sec:shellmodel}

We  employ a well-known Gledzer-Ohkitani-Yamada (GOY) shell model\cite{Yamada:PRE1998,Gledzer:DANS1973} of hydrodynamic and  MHD turbulence  to quantify  the kinetic energy flux, root mean square (rms) velocity, and  nonlinear term connected to $ {\bf (u \cdot \nabla) u}$   discussed  in previous sections.  Based on these quantities we will show that the turbulent drag is indeed reduced in MHD turbulence compared to hydrodynamic turbulence.
	
	The  equation for the GOY shell model for hydrodynamic turbulence is 
\be
\frac{du_n}{dt}=N_n[u,u]-\nu k_n^2u_n+F_{\mathrm{ext},n},\label{shell_velo}
\ee
where $u_n$  and $F_{\mathrm{ext},n}$ represent respectively the velocity and external force fields for shell $n$, $k_n = k_0 \lambda^n$ is the wavenumber of the $n$th shell, and $\nu$ is the kinematic viscosity. Note that  $\lambda$ is a constant, and it is taken to be golden  ratio~\cite{Lvov:PRE1998}. The mathematical expression of $N_u[u,u]$ is 
\bea
N_n[u,u]&=&-i(a_1 k_n u^*_{n+1}u^*_{n+2} + a_2 k_{n-1} u^*_{n+1}u^*_{n-1}\nonumber\\ &&
+ a_3 k_{n-2} u^*_{n-1}u^*_{n-2}),\label{eq:nlin_shell_model}
\eea
where $a_1$, $a_2$ and $a_3$ are constants. The conservation of kinetic energy ($\int d\mathbf{r}(u^2 /2)$) and kinetic helicity ($\int d\mathbf{r} (\bf{u} \cdot \boldsymbol{\omega})$, where $\boldsymbol{\omega}$ denotes vorticity field) for force-less and dissipationless regime helps us determine the values of  constant $a_1, a_2, a_3$.  A choice of the constants used in previous simulations and also in this paper are $a_{1}=\lambda$, $a_{2}=1-\lambda$ and $a_{3}=-1$.  

There are many shell models for MHD turbulence, e.g., see\cite{Frick:PRE1998, Stepanov:JoT2006,  Plunian:PR2012}.  However, in this paper we employ the GOY-based shell model  proposed by Verma and Kumar\cite{Verma:JoT2016}, which is 
\bea
\frac{du_n}{dt}&=&N_n[u,u]+M_n[B,B]-\nu k_n^2u_n+F_{\mathrm{ext},n}, \label{velo_eqn_MHD}\\
     \frac{dB_n}{dt}&=&O_n[u,B]+P_n[B,u]-\eta k_n^2B_n,\label{mag_field_eqn_MHD}
\eea
where $B_{n}$ is the shell variable for the magnetic field, and $\eta$ is the magnetic diffusivity. In addition to $N_u[u,u]$ of Eq.~(\ref{eq:nlin_shell_model}), there are three additional nonlinear terms $M_n[B,B]$, $O_n[u,B]$ and $P_n[B,u]$, which are 
\bea
M_n[B,B]&=&-2i(b_1 k_n B^*_{n+1}B^*_{n+2} + b_2 k_{n-1} B^*_{n+1}B^*_{n-1}\nonumber\\ && + b_3 k_{n-2} B^*_{n-1}B^*_{n-2}),\\
O_n[u,B]&=& -i[k_n ( d_1 u^*_{n+1}B^*_{n+2} + d_3 B^*_{n+1} u^*_{n+2})\nonumber \\ && 
    + k_{n-1} ( -d_3 u^*_{n+1}B^*_{n-1} + d_2 B^*_{n+1} u^*_{n-1}) \nonumber\\ &&
    - k_{n-2} ( d_1 u^*_{n-1}B^*_{n-2} + d_2 B^*_{n-1} u^*_{n-2})],\\
P_n[B,u]&=& i[k_n ( b_2 u^*_{n+1}B^*_{n+2} + b_3 B^*_{n+1} u^*_{n+2}) \nonumber \\ && 
    + k_{n-1} ( b_3 u^*_{n+1}B^*_{n-1} + b_1    B^*_{n+1} u^*_{n-1}) \nonumber \\ &&
    + k_{n-2} ( b_2 u^*_{n-1}B^*_{n-2} + b_1 B^*_{n-1} u^*_{n-2})],
\eea
where $\{b_1, b_2, b_3\}$ and $\{d_1$, $d_2$, $d_3\}$ are the constants. Akin to  hydrodynamic turbulence, these constants are  computed using the conservation of the total energy ($\int d\mathbf{r} (u^2 + B^2) /2$), the magnetic helicity ($\int d\mathbf{r} (\mathbf{A} \cdot \mathbf{B})$, where $\mathbf{A}$ is magnetic vector potential) and the cross helicity ($\int d\mathbf{r} (\bf {u} \cdot \bf{B})$) for $\nu=\eta=0$ and $F_{\mathrm{ext},n}=0$.  It is also ensured that the pure fluid case ($B_{n} =0$) satisfies the conservation of kinetic energy and kinetic helicity. These constraints yield the following values of the constants~\cite{Verma:JoT2016}:
\bea
b_1=\lambda, b_2 = 1+\frac{\lambda}{2}, b_3 = -1-\frac{3\lambda}{2}, \\
d_1 = \frac{5\lambda}{2}, d_2 = -\lambda + 2, d_3 = -\frac{\lambda}{2}.
\eea
Note that the expression of $N_n[u,u]$ and the constants $a_1$, $a_2$, and $a_3$ are same as those for hydrodynamic turbulence discussed earlier.   

We simulate the shell models for hydrodynamic and MHD turbulence for identical kinetic energy injection rate, and compare the kinetic energy fluxes, $\la | ({\bf u}\cdot \nabla) {\bf u} | \ra$,  and the rms velocities for the two models.  For the simulations, we divide wavenumbers  into $36$ logarithmically binned shells and take $\lambda = (\sqrt{5} +1 )/2$, the golden ratio.   For time integration, we use Runge-Kutta fourth order (RK4) scheme with a fixed $\Delta t$.   We carry out our hydrodynamic and MHD simulation up to  1000 eddy turnover time unit.

As illustrated in Figures 1 and 2, we force the large-scale velocity field using  external random force at shells $n = 1$ and $2$ such that the kinetic energy supply rate is maintained at a constant value.  For the same, we employ the scheme proposed by Stepanov and Plunian~\cite{Stepanov:JoT2006}.   We perform three sets of simulations with  kinetic energy supply rates $\epsilon_\mathrm{inj}=0.1, 1.0$ and $10.0$, and  $\nu,  \eta=10^{-6}$.  For  $\epsilon_\mathrm{inj}=0.1$ and $1.0$, we choose $\Delta t =5\times10^{-5}$,  but  for $\epsilon_\mathrm{inj}=10.0$, we take $\Delta t =10^{-5}$.  The numerical results are summarized in Table~\ref{tab:HD_MHD}.

\begin{figure}%[tbhp]
	\centering
	\includegraphics[width=1\linewidth]{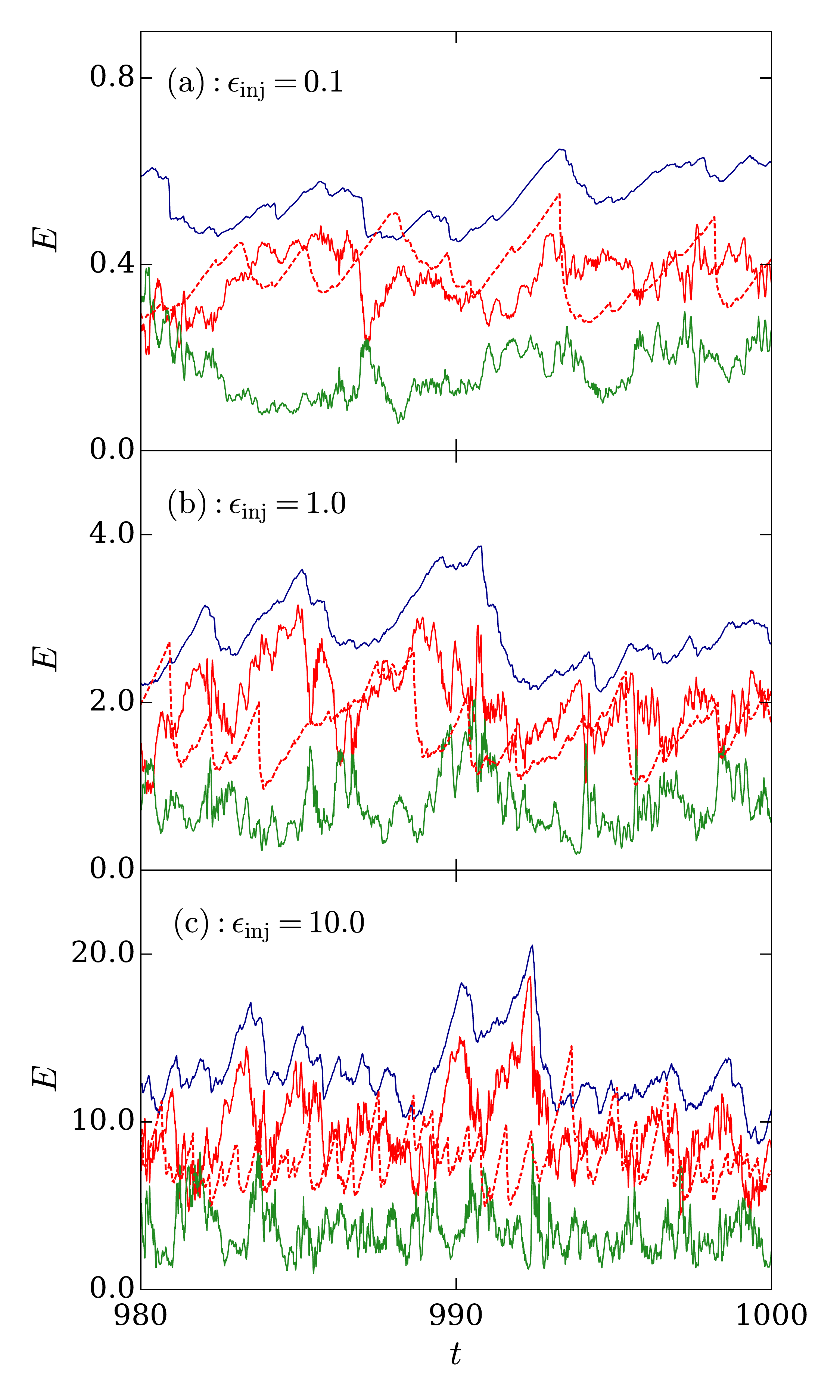}
	\caption{(color online) Time evolution plots of the kinetic energy for hydrodynamics (red dashed curves), as well as those of the kinetic energy (solid red curves), magnetic energy (green curves), and total energy (dark-blue curves) for MHD turbulence.  The three subplots represent the results for  injection rates (a) $\epsilon_\mathrm{inj}=0.1$, (b) $\epsilon_\mathrm{inj} = 1.0$, and (c) $ \epsilon_\mathrm{inj} = 10.0$. }
	\label{fig:energy_comp}
\end{figure}

The simulations reach their respective steady states after around $200$ eddy turn over times.  In Figure~\ref{fig:energy_comp}, for the steady states, we plot the time series of the kinetic energy for hydrodynamic simulations;  and kinetic, magnetic, and total energies for MHD simulations for the three injection rates. The time series demonstrate that for the same $\epsilon_\mathrm{inj}$,  the kinetic energy for MHD turbulence is typically larger than that for the hydrodynamic turbulence.  These observations clearly demonstrate an enhancement of the speed in MHD turbulence than hydrodynamic turbulence, as described in Sec.~\ref{sec:MHD}.

\begin{table}%[tbhp]
	\caption{For the three sets of shell model simulations of hydrodynamic and MHD turbulence with  kinetic energy injection rates $\epsilon_\mathrm{inj}=0.1,1.0$ and $10.0$, the table contains the numerical values of kinetic energy flux ($\Pi_{u}$) in the inertial range;  rms velocity ($U$); and $\la | ({\bf u}\cdot \nabla) {\bf u} | \ra= \la \sum_n |N_n[u,u]|^2  \ra^{1/2}$. }
	\vspace{10pt}
	\begin{tabular}{c|c|c|c|c|c|c}
		\hline
		\hline
		{}&\multicolumn{3}{|c}{Hydrodynamics}&\multicolumn{3}{|c}{MHD}\\
		\hline
		$\epsilon_\mathrm{inj}$ & $\Pi_{u}$ & $U$ & $\la | ({\bf u}\cdot \nabla) {\bf u} | \ra$ & $\Pi_{u}$ & $U$ & $\la | ({\bf u}\cdot \nabla) {\bf u} | \ra$ \\
		\hline
		$0.1$ & $0.1$ & $0.87$ & $8.77$ & $0.02$ & $0.92$ & $4.17$\\
		\hline
		$1.0$ & $1.0$ & $1.88$ & $47.48$ & $0.21$ & $2.02$ & $23.79$\\
		\hline 
		$10.0$ & $10.0$ & $3.95$ & $271.88$ & $2.06$ & $4.33$ & $136.44$\\
		\hline
	\end{tabular}
	\label{tab:HD_MHD}
\end{table}

For hydrodynamic and MHD turbulence, using the numerical $u_n$ from the shell models, we estimate the rms value of the velocity using
\be
U
= \la  \sum_n |u_n|^2  \ra^{1/2},
\ee
 and the rms values of $({\bf  u}\cdot \nabla) {\bf u}$  using the formula:
\be
\la | ({\bf u}\cdot \nabla) {\bf u} | \ra 
= \la \sum_n |N_n[u,u]|^2  \ra^{1/2}.
\ee
We average the computed values over 50000 frames during the steady state.  These values are further averaged over 16 different simulations that were started with random initial condition.   The results   listed in Table~\ref{tab:HD_MHD} indicate  a relative suppression of $\la | ({\bf u}\cdot \nabla) {\bf u} | \ra $  in  MHD turbulence compared to hydrodynamic turbulence. However 
\be
U_\mathrm{MHD} > U_\mathrm{HD} 
\ee
indicating a relative enhancement of the velocity in MHD turbulence.  Note that the $\la | ({\bf u}\cdot \nabla) {\bf u} | \ra $ depends critically on the phases of the Fourier modes; larger $U$ does not necessarily imply larger $\la | ({\bf u}\cdot \nabla) {\bf u} | \ra$.

\begin{figure}%[tbhp]
	\centering
	\includegraphics[width=1\linewidth]{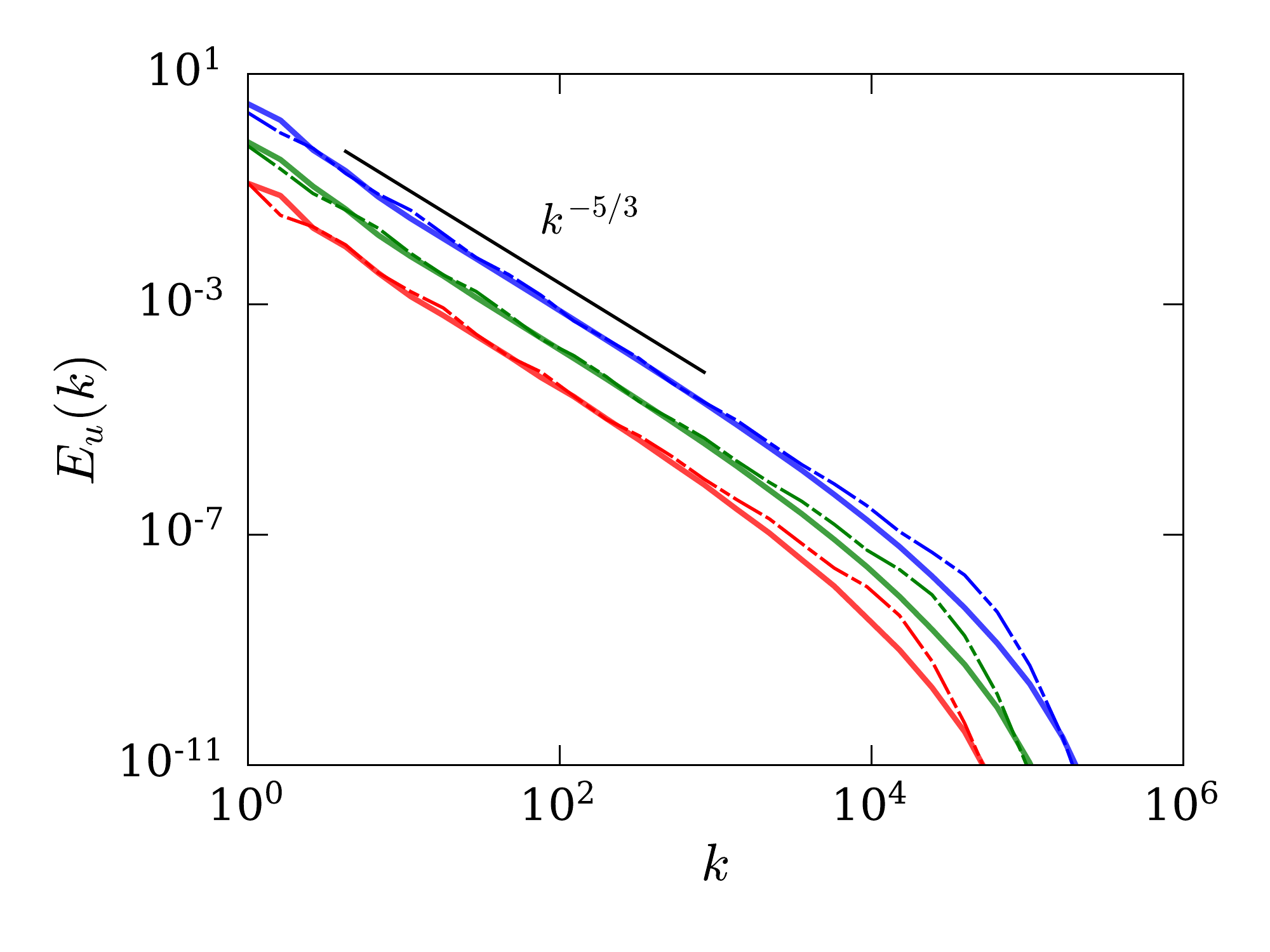}
	\caption{  (color online) Plots of kinetic energy spectra $E_u(k)$ for injection rates $\epsilon_\mathrm{inj}=0.1$ (red), $\epsilon_\mathrm{inj}= 1.0$ (green), and $\epsilon_\mathrm{inj}= 10.0$ (blue). The solid curves represent the $E_u(k)$ for MHD turbulence, and the dashed curves for hydrodynamic turbulence. Kolmogorov's $-5/3$ scaling (black) fits well in the inertial range for all the cases.}
	\label{fig:spectrum_comp}
\end{figure}

 Using the numerical data, we also compute the averaged kinetic energy spectrum 
 \be
 E_u(k) = \frac{1}{2 k_n} \la {u_n^2} \ra 
 \ee
 for hydrodynamic and MHD turbulence.  We adopt the same averaging procedure as done for the computation of $U$ and $\la | ({\bf u}\cdot \nabla) {\bf u} | \ra $.  The energy spectra exhibited in Figure~\ref{fig:spectrum_comp} for the three sets show $k^{-5/3}$ power law in the inertial range.  For each case, $ E_u(k)$ for hydrodynamic and MHD turbulence are almost equal to each other, except at small and large wavenumbers.  At small wavenumbers  $E_u(k)$ for MHD turbulence is larger than that for hydrodynamics, consistent with the  observation that $U_\mathrm{MHD} > U_\mathrm{HD}$.     Also, note that the trend is reversed at large wavenumbers,  whose role in turbulent drag reduction needs to be explored in future.

\begin{figure}%[tbhp]
	\centering
	\includegraphics[width=1\linewidth]{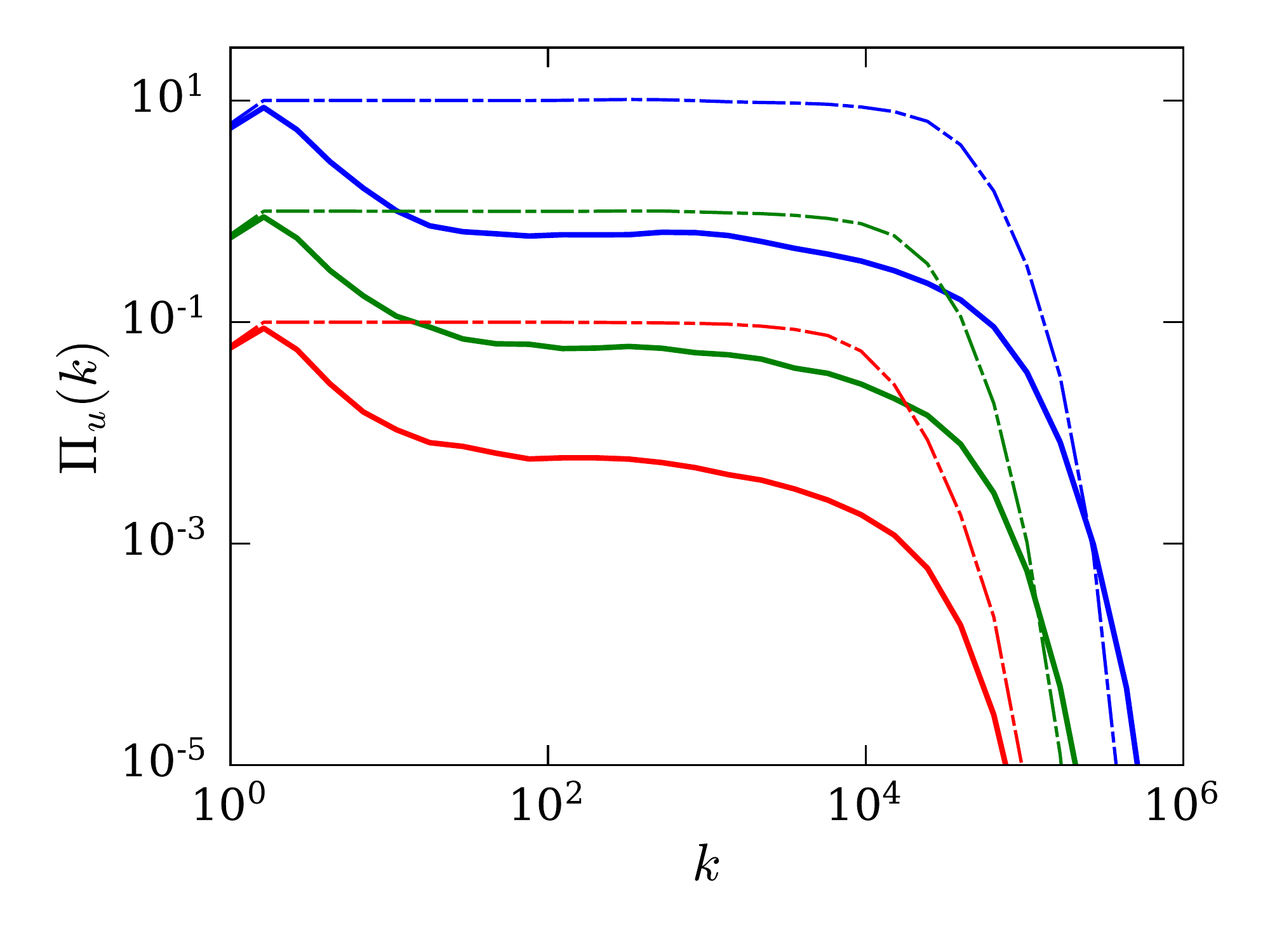}
	\caption{ (color online) Plots of kinetic energy fluxes $\Pi_u(k)$ for the different injection rates $\epsilon_\mathrm{inj}=0.1$ (red), $\epsilon_\mathrm{inj}= 1.0$ (green), and $\epsilon_\mathrm{inj}= 10.0$ (blue). The solid curves represent the $\Pi_u(k)$ for MHD turbulence, while the dashed curves for hydrodynamic turbulence.}
	\label{fig:flux_comp}
\end{figure}

Lastly, we compute  averaged kinetic energy fluxes for hydrodynamic and MHD turbulence using the following flux formula~\cite{Verma:JoT2016, Verma:book:ET}:
\bea
\Pi_{u}(K) & = & a_3 k_{K-1} \la \Im(u_{K-1}^* u_K^* u_{K+1}^*) \ra \nonumber \\
& &-  a_1 k_{K} \la \Im(u_{K}^* u_{K+1}^* u_{K+2}^*) \ra  .
\eea
In Figure~\ref{fig:flux_comp}, we plot the kinetic energy fluxes for the three $\epsilon_\mathrm{inj}$.  The figure   clearly show that for all the three cases,
\be
\Pi_{u,\mathrm{MHD}} < \Pi_{u,\mathrm{HD}}.
\ee
 We also compute $\Pi_B$ as $\epsilon_\mathrm{inj} - \Pi_u$, and observe it to be positive.  That is, the magnetic field receives energy from the velocity field.  These results are  consistent with the discussion of Sec.~\ref{sec:MHD}.

 The above results on $\Pi_{u}(K)$ are  consistent with the fact that $\la | ({\bf u}\cdot \nabla) {\bf u} | \ra$ for MHD turbulence is lower than the corresponding term for hydrodynamic turbulence (see Table~\ref{tab:HD_MHD}); lower $\la | ({\bf u}\cdot \nabla) {\bf u} | \ra$  leads to lower kinetic energy flux. In addition, lower $\la | ({\bf u}\cdot \nabla) {\bf u} | \ra$ produces larger $U$ for MHD turbulence.    We also remark that the  above trends on the energy fluxes are  consistent with the Mininni et al.'s results \citep{Mininni:ApJ2005} based on direct numerical simulations. 

Thus, our numerical simulations demonstrate that MHD turbulence has  lower $\la | ({\bf u}\cdot \nabla) {\bf u} | \ra$ and  kinetic energy flux, but a larger $U$ compared to hydrodynamic turbulence.  These results demonstrate that turbulent drag is indeed reduced in MHD turbulence.

 In the next section we will describe turbulent drag reduction in QSMHD turbulence.

\section{Drag reduction in QSMHD turbulence in terms of energy flux}
\label{sec:QSMHD}

Quasi-static MHD turbulence, a special class of MHD flows, has very small magnetic Prandtl number~\cite{Knaepen:ARFM2008,Verma:ROPP2017}.   A typical example of QSMHD is a  liquid metal flow with a strong external magnetic field. In a nondimensionalized set of equations, the Lorentz force is proportional to $-N {\bf u }$, hence it is dissipative~\cite{Knaepen:ARFM2008,Verma:ROPP2017}.  The parameter $N$, called {\em interaction parameter}, is the ratio of the Lorentz force and nonlinear term ${\bf (u \cdot  \nabla) u}$, or
\be
N = \frac{\sigma B_0^2 L}{\rho U},
\ee
where $\sigma $ { is} the electrical conductivity, and ${\bf B}_0$ is the external magnetic field, which is assumed to be constant.

  In QSMHD turbulence~\cite{Knaepen:ARFM2008,Verma:ROPP2017},  the corresponding force and related energy transfer rates are
 \bea
 {\bf F}_u({\bf k})  & =  &  -N {\bf u(k)}  \cos^2 \theta , \\
  \mathcal{F}_u({\bf k}) & = &  \Re[ {\bf F}_u({\bf k}) \cdot {\bf u^*(k)}] = - 2 N  E_u({\bf k})  \cos^2 \theta<0 , \nonumber \\
 \eea
where  $\theta$ is the angle between the external magnetic field ${\bf B}_0$ and the wavenumber ${\bf k}$.   Therefore,  the energy transferred from the velocity field  to the magnetic field, $ \Pi_B(K)$, takes the following form:
 \be
 \Pi_B(K)=-\sum_{k\le K} \mathcal{F}_u({\bf k}) = \sum_{k\le K}  2 N   E_u({\bf k})  \cos^2 \theta>0.
 \ee
 Hence, the  dissipative  Lorentz force transfers the kinetic energy to the magnetic energy, which is immediately destroyed by Joule dissipation.   
 Hence, as argued above, the kinetic energy flux is suppressed compared to the hydrodynamic turbulence.

As argued in the previous section, the depleted $\Pi_u$ in QSMHD turbulence  leads to a reduction in turbulent drag.  Physically, the magnetic field  smoothens the velocity field compared to hydrodynamic turbulence.  Therefore, we expect the turbulent drag  in QSMHD turbulence to be lower than the corresponding hydrodynamic counterpart.

Reddy and Verma~\cite{Reddy:PF2014} simulated QSMHD turbulence for a wide range of interaction parameter $N$ with a constant  energy injection rate of 0.1 (in nondimensional unit).  In Table~\ref{tab:QSMHD} we list  the rms velocity $U$ as a function of $N$.   Here, $U$ is measured in units of $L/T$, where $L,T$ are length and time scales of the large scale eddies.  Clearly,  $U$ increases monotonically with $N$.  In other words,  { the turbulent drag is reduced in QS MHD turbulence.}   It is important to note that a large $U$ does not necessarily imply a large nonlinear term ${\bf (u \cdot \nabla) u}$, which  depends on $U$, as well as on the phase relations between the velocity modes.  The magnetic field alters the phase relations in ${\bf (u \cdot \nabla) u}$; thus suppresses this term and the drag; and produces larger $U$.

 \begin{table}%[tbhp]
\caption{For simulation of QSMHD turbulence by Verma and Reddy~\cite{Verma:PF2015QSMHD}, root mean square (rms) velocity of the flow ($U$) as a function of interaction parameter $N$.  The flow speed increases with the increase of $N$. }
\begin{tabular}{c|cccc}
\hline
$N$ & 1.7 & 18 & 27 & 220  \\  \hline
$U$ & 0.39 & 0.51 & 0.65 & 0.87	 \\ \hline
\end{tabular}
\label{tab:QSMHD}
\end{table}

The reduced turbulent flux plays an important role in drag reduction.   Note however that such reduction does not occur in laminar QSMHD because the Lorentz force damps the flow further.   For example, in a QSMHD channel flow, the maximum velocity in the centre of the channel is~\citep{Muller:book, Verma:ROPP2017}
\be
U_\mathrm{QSMHD} =- \frac{1}{\sigma B_0^2} \left( \frac{\partial p}{\partial x} \right).
\ee
In contrast,  in a hydrodynamic channel flow, the maximum velocity is~\citep{Kundu:book}
\be
U_\mathrm{HD} =- \frac{d^2}{2 \nu \rho} \left( \frac{\partial p}{\partial x} \right),
\ee
where $d$ is half-width of the channel.  The ratio of the two velocities is
\be
\frac{U_\mathrm{QSMHD}}{U_\mathrm{HD} } = \frac{2 \nu \rho}{\sigma B_0^2 d^2} = \frac{1}{\mathrm{Ha}^2},
\ee
where $\mathrm{Ha}$ is the Hartmann number, which is much larger than unity for QSMHD turbulence.  Hence, the velocity in laminar QSMHD is much smaller than that in hydrodynamic channel.  In contrast, in a turbulent flow, the Lorentz force suppresses $\la | ({\bf u}\cdot \nabla) {\bf u} | \ra$, and hence increases the mean velocity.   Thus, the aforementioned drag reduction is a nonlinear phenomena, which ia related to the energy flux.   {We remark that such reduction in fluctuations have been observed in laboratory experiments.  For example, \citet{Berhanu:PRE2008} observed significant reduction in velocity fluctuations in a turbulent flow of gallium under  an external magnetic field. In addition, suppression of fluctuations in QSMHD turbulence finds applications in engineering, for example, in crystal growth and plate rolling\cite{Bojarevics:book_chapter, Bochkarev:book_chapter, Chudinovskij:book_chapter}. }

In the next section, we show that a similar process is at work in turbulent flows with dilute polymers.  We discuss this briefly to make a connection between turbulent drag reduction in MHD turbulence with that in a polymeric solution.
 
%%%% 
\section{Revisiting turbulent drag reduction with dilute polymers}
\label{sec:polymer}

Among a large body of works on turbulent drag reduction in polymers\cite{Tabor:EPL1986,deGennes:book:Intro,Sreenivasan:JFM2000,Benzi:PD2010,Benzi:ARCMP2018},  those related to the energy fluxes of polymeric flows are quite small in number.   Recently, Valente et al.\citep{Valente:JFM2014,Valente:PF2016} performed numerical simulations of polymeric solution and computed various energy fluxes.  They showed a transfer of kinetic energy to the elastic energy of polymers for a set of parameters.    \citet{Nguyen:PRF2016} reported similar energy transfers, but they also showed a transfer from the elastic energy to the kinetic energy at small scales.  Using numerical simulations, Benzi et al.~\citep{Benzi:PRE2003} and Perlekar et al.~\citep{Perlekar:PRL2006}  analysed the energy spectra and dissipation rates of kinetic and elastic energies.  \citet{Thais:IJHFF2013} computed the above quantities using turbulent stresses.  \citet{Ray:EPL2016}, and \citet{Kalelkar:PRE2005} studied the turbulent drag reduction in polymers using shell model.   Several numerical simulations explore the effects of various parameters---elasticity, Reynolds number, geometry of polymers---on turbulent drag reduction~\citep{Boffetta:PRE2005, Benzi:PRE2008}.   In this paper we  relate the energy fluxes in polymeric turbulence to turbulent drag reduction.

Polymers are often described using  finitely extensible nonlinear elastic-Peterlin (FENE-P) model~\citep{Sagaut:book, Benzi:PD2010}.  In this model, the equations for a turbulent flow with dilute polymer in tensorial form are~\citep{Sagaut:book, Benzi:PD2010, Fouxon:PF2003}
\bea
\frac{\partial{u_i}}{\partial t} +u_j \partial_j u_i
& = & -\partial_i p /\rho +  \nu \partial_{jj} u_i + \frac{\mu}{\tau_p} \partial_j (f   w_{ij})
 + F_{\mathrm{ext}, i}, \label{eq:tensor:FENE-u} \nonumber \\ \\
\frac{\partial{w_{ij}}}{\partial t} +  u_l \partial_l w_{ij}
& = &     w_{il} \partial_l u_j +  w_{jl} \partial_l u_i +  \frac{1}{\tau_p}   [f w_{ij} -\delta_{ij} ]
\label{eq:tensor:FENE-C} \\
\partial_i u_i  & = & 0, 
\eea
\label{eq:tensor:FENE}
 where $\rho$ is the mean density of the solvent, $\nu$ is the kinematic viscosity, $\mu$ is an additional viscosity parameter, $\tau_p$ is the polymer relaxation time, and $f$ is the renormalized Peterlin's function.    In the above equations, the following forces are associated with ${\bf u}$ and $w_{ij}$ (apart from constants):
 \bea
 F_{u,i} & = &   \partial_j (f   w_{ij}), \\
 F_{w,ij} & = &w_{il} \partial_l u_j + w_{jl} \partial_l u_i.
 \eea
In Fourier space, the energy feed to kinetic energy by ${\bf F}_u$ is
  \bea
 \mathcal{F}_u({\bf k}) & = &  - \sum_{\bf p} \Im \left[  k_j f({\bf q}) w_{ij}({\bf p}) u^*_i({\bf k}) \right], 
 \eea
where ${\bf q = k-p}$.   The net energy transferred  from the velocity modes inside the sphere of radius $K$ to all the modes of the polymer is
\be
 \Pi_w(K) = \sum_{k \le K}   -\mathcal{F}_u({\bf k}).
 \ee

Using numerical simulations of  polymeric turbulence, \citet{Valente:JFM2014,Valente:PF2016} analysed the energy transfers, in particular, the fluxes $\Pi_u$ and $\Pi_w$.  These fluxes depend on the Deborah number, $\mathrm{De}$,  which is the ratio of the relaxation  time scale of the polymer and the characteristic time scale for the energy cascade.  A common feature among all the numerical runs  is that  $\Pi_w >0$.  In addition, the  transfer from kinetic energy to elastic energy is maximum when $\mathrm{De} \sim 1$. For example, Valente et al.~\cite{Valente:JFM2014} showed that for $\mathrm{De} = 1.17$, $\Pi_w/\epsilon_\mathrm{inj} \approx 0.8$ for $k\eta > 0.2$ where $\eta$ is Kolmogorov's wavenumber; and $\Pi_u/\epsilon_\mathrm{inj}$ peaks to approximately 0.3 near $k\eta \approx 0.1$.  Thus, $\Pi_u$ is  reduced to around 20\% to 30\% for this case.  We expect similar reduction for other cases when $\mathrm{De} \sim 1$.  Using a shell model, \citet{Ray:EPL2016} showed that the kinetic energy is  transferred to the polymer elastic energy.  \citet{Thais:IJHFF2013}, and \citet{Nguyen:PRF2016} arrived at  similar conclusions based on their direct numerical simulation of polymeric turbulence.  Based on these observations, we deduce that
 \be
  \Pi_{u,\mathrm{Polymeric}} <  \Pi_{u,\mathrm{HD}}.
  \ee   
 This reduction leads to a decrease in $\la | ({\bf u}\cdot \nabla) {\bf u} | \ra$ and  in turbulent drag (also see Sec.~\ref{sec:MHD}).  Note however that in the laminar regime, the energy flux vanishes  and hence the drag is not reduced.  This is consistent with the observations by \citet{Sreenivasan:JFM2000}, and  with the formulas for laminar QSMHD discussed in the previous section.  It will be interesting  to perform a  comparative study between the shell models for polymers and hydrodynamic, as in Sec.~\ref{sec:shellmodel}.

Earlier, Fouxon and Lebedev~\citep{Fouxon:PF2003} showed that the equations for dilute polymers are intimately connected to those of MHD turbulence.  Hence, the aforementioned energy transfers~\cite{Valente:JFM2014, Ray:EPL2016,Thais:IJHFF2013} from the kinetic energy  to elastic energy is consistent with the results of MHD turbulence.   

We end this section with a cautionary remark.  Turbulent drag reduction in a polymer solution is attributed to many factors: boundary layers, viscoelasticity, turbulence in the bulk, interactions between polymers and large scale velocity field, anisotropy, polymer concentration, elasticity parameters,    Reynolds number, etc.~\citep{Tabor:EPL1986,deGennes:book:Intro,Sreenivasan:JFM2000,Benzi:PD2010,Benzi:ARCMP2018, Boffetta:PRE2005, Benzi:PRE2008}  In the paper we focus on  the reduction in kinetic energy flux due to polymers.  More detailed and  comprehensive simulations and experiments with all the above parameters are required for a definitive conclusion.  

Turbulent drag reduction has also been reported in flows with bubbles~(e.g.~\citet{Lvov:PRL2005}).   It has been argued that kinetic energy may be transferred to the elastic energy of the bubbles; this process may be important for the turbulent drag reduction in bubbly turbulence.  

We conclude in the next section.
 
\section{Discussions and conclusions}
\label{sec:conclusions}

In this paper we show that MHD turbulence  and  quasi-static MHD (QSMHD) turbulence exhibit turbulent drag reduction.  We relate the above to the reduced kinetic energy flux due to a  partial transfer of kinetic energy flux to the magnetic field.  This process  leads to a suppression of the nonlinear term $\la | ({\bf u}\cdot \nabla) {\bf u} | \ra $, and {stronger rms  velocity $ U $}  compared to hydrodynamic turbulence.     
Similar reduction in kinetic energy flux has been reported in solutions with dilute polymers and bubbles.  Thus,   the energy flux provides a useful perspective on turbulent drag reduction. {We remark that some of the  works by Eyink et al.\cite{Eyink:ApJ2011, Eyink:Nature2013} and	Jafari et al.\cite{Jafari:PRE2019, Jafari:PRE2019_diffusion, Jafari:PRE2020} are useful in quantifying randomness  in hydrodynamic and MHD turbulence.  These authors also studied stochasticity  in dynamo. }

 We simulated shell models of hydrodynamic and MHD turbulence, and computed the rms velocities, $\la | ({\bf u}\cdot \nabla) {\bf u} | \ra $, and kinetic energy fluxes for the two cases.  For the same kinetic energy injection rate,  $\la | ({\bf u}\cdot \nabla) {\bf u} | \ra $ and kinetic energy flux in MHD turbulence are lower than those for hydrodynamic turbulence.  However, the rms velocity is larger for MHD turbulence; this enhanced velocity still yields  suppression in $\la | ({\bf u}\cdot \nabla) {\bf u} | \ra $ in MHD turbulence due to the phase relations of the complex Fourier modes.  Our simulations also show that in MHD turbulence, a finite amount of energy  is transferred from the velocity field to the magnetic field.  These numerical observations demonstrate turbulence drag reduction in MHD turbulence.  Direct numerical simulations of hydrodynamic and MHD turbulence will provide stronger credence to the above theory.  However, such simulations are very expensive, and they are planned for future.
	
%It has been reported in MHD turbulence literature that a strong mean magnetic field too suppresses turbulence\cite{Oughton:JFM1994,Sundar:PP2017}.  It occurs due to two-dimensionalization of the flow.  Hence, the mechanisms for turbulence suppression by a mean field   and by a random magnetic field differ significantly.

Turbulent drag reduction in MHD turbulence has important ramifications in astrophysical and engineering flows.   We believe that this feature would have consequences in solar and planetary dynamos.   In a subcritical dynamo transition, a finite magnetic field appears at one transition point, and a finite magnetic field shutdown abruptly at the other transition point\cite{Yadav:PRE2012,Verma:PP2013}.  When we carry forward our results discussed in this paper to such a  system, we expect that the generated magnetic field would suppress the nonlinear term $\la | ({\bf u}\cdot \nabla) {\bf u} | \ra$ and enhance $U$, while shutdown of the dynamo would enhance $\la | ({\bf u}\cdot \nabla) {\bf u} | \ra$ and suppress $U$.    \citet{Yadav:PRE2012} observed precisely this feature in their simulation of subcritical dynamo transition with small magnetic Prandtl number.   We can relate this result to the Martian dynamo.  At present, Mars has no magnetic field, but there are  evidences of strong crustal magnetic field.  Hence, researchers believe that  dynamo action was present in Mars in the past, but it got shutdown via a  subcritical dynamo transition~\cite{Kuang:GRL2008}.   Based on the above arguments, we can conjecture that during this dynamo shutdown,  $\la | ({\bf u}\cdot \nabla) {\bf u} | \ra$ may have increased, while the  mean  $U$ may have been weakened.  We hope that numerical simulations may verify this conjecture.

In addition, we expect that our findings would be useful for understanding the dynamics of some of the laboratory dynamos employed to understand geodynamo\cite{Lathrop:PT2011}. Note that the large laboratory devices for studying dynamo transitions need to be robust enough to handle the aforementioned sudden increase in the large-scale velocity field. These issues would be of concern to the designers of the experiments.  In addition, the magnetic field can be used to suppress fluctuations or the nonlinear term $ {\bf u \cdot \nabla u} $ in engineering flows,  as in  crystal growth and plate rolling\cite{Bojarevics:book_chapter, Bochkarev:book_chapter, Chudinovskij:book_chapter}.   These ideas help in producing smooth plates for aerodynamic designs and  defect-free large crystals.

%For a conclusive understanding of drag reduction, we need a detailed study of the flow including the bulk and boundary layers.  In addition, effects of wide range of parameters (polymer concentration, Deborah number, Reynolds number) on turbulent drag reduction need to be explored.   Also, for a more quantitative description, it is important to compute $c_1, c_2$ of Eqs.~(\ref{eq:c1}, \ref{eq:c2}) ffAor the  flows discussed in this paper.  The connection between the energy flux and turbulent drag  presented in this paper is an important link in this puzzle, but more comprehensive works  are required for a definitive conclusion.

Turbulent drag reduction for aircrafts and automobiles is an interesting and challenging problem.  Towards this objective, researchers have devised many interesting schemes including laminar flow control, control of boundary layer detachment, injection of travelling waves, etc.~\citep{Bushnell:PIME2003,Mamori:PF2014} This phenomena is related to the relaminarization of fluid flows.  \citet{Narasimha:AAM1979} studied relaminarization in stably stratified turbulence, rotating turbulence, and accelerated flows, and connected relaminarization to the reduction in the nonlinear term $\la | ({\bf u}\cdot \nabla) {\bf u} | \ra $.  Connecting the above phenomena to the energy flux would be an interesting and useful exercise. 

In summary, turbulent energy flux provides valuable insights into the dynamics of drag reduction.

 %\showmatmethods{} % Display the Materials and Methods section

\acknowledgements
The authors thank Abhishek Kumar, Franck Plunian, K. R. Sreenivasan, Shashwat Bhattacharya, and Supratik Banerjee for useful  discussions.   Soumyadeep Chatterjee is supported by INSPIRE fellowship (IF180094) of Department of Science \& Technology, India. 

%\nocite{apsrev41Control}
%\bibliographystyle{ieeetr_abhi}
%\bibliographystyle{apsrev4-1}
%\bibliographystyle{act}
%\bibliography{/Users/mkv/Dropbox/docs-pub/bib/journal,/Users/mkv/Dropbox/docs-pub/bib/book,/Users/mkv/Dropbox/docs-pub/bib/book_chapter,/Users/mkv/Dropbox/docs-pub/bib/preprint}


\end{document}